# Development of Personalized Sleep Induction System based on Mental States

Young-Seok Kweon
*Dept. Brain and Cognitive Engineering*
Korea University
Seoul, Republic of Korea
youngseokkweon@korea.ac.kr

Gi-Hwan Shin
*Dept. Brain and Cognitive Engineering*
Korea University
Seoul, Republic of Korea
gh_shin@korea.ac.kr

Heon-Gyu Kwak
*Dept. Artificial Intelligence*
Korea University
Seoul, Republic of Korea
hg_kwak@korea.ac.kr

*Abstract*—Sleep is an essential behavior to prevent the decrement of cognitive, motor, and emotional performance and various diseases. However, it is not easy to fall asleep when people want to sleep. There are various sleep-disturbing factors such as the COVID-19 situation, noise from outside, and light during the night. We aim to develop a personalized sleep induction system based on mental states using electroencephalogram and auditory stimulation. Our system analyzes users' mental states using an electroencephalogram and results of the Pittsburgh sleep quality index and Brunel mood scale. According to mental states, the system plays sleep induction sound among five auditory stimulation: white noise, repetitive beep sounds, rainy sound, binaural beat, and sham sound. Finally, the sleep-inducing system classified the sleep stage of participants with 94.7% and stop auditory stimulation if participants showed non-rapid eye movement sleep. Our system makes 18 participants fall asleep among 20 participants.

## I. INTRODUCTION

Sleep is a clearly necessary process to keep our life healthy although there are many mysteries about sleep [1], [2]. A reduction of cognitive ability, depression, and impairment of motor function has happened when people poorly sleep [3], [4]. Sleep deprivation is also being highlighted as a cause of metabolic diseases such as diabetes and obesity [5]. One of the main causes of sleep deprivation is the difficulty of sleeping at the right time [6]. This is because there are various disturbing factors such as negative emotion, the COVID-19 situation, noise from outside, and light during the night [7]–[9]. Some disturbing factors can be easily removed in a simple way. However, emotional and cognitive factors are difficult to handle because they interact with a person complicatedly.

A person who had a problem inducing sleep has an interest in auditory stimulation like music, white noise (WN), binaural beat (BB), etc. The BB induces oscillation of frequency difference to the brain by delivering oscillation at two adjacent frequencies to each ear at the same time [10], [11]. Especially, a recent study showed the possibility of sleep induction by 6Hz BB combined with natural sounds [12]. Unlike BB, WN uses all different frequencies with the same power. WN's sleep induction effects have been evaluated for neonates and patients in intensive and coronary care units [13]–[15]. They fell asleep better when WN was given than WN was not given. We sometimes experience micro-sleep during monotonous tasks and repetitive beep sounds also induce sleep [16], [17]. Although WN, BB, and RB are artificial auditory sounds, natural sounds are also popular to induce sleep on the internet. Interestingly, 15 million views were reported on YouTube about rainy sound (RS). However, there is no magical lullaby that makes everyone fall asleep. Since every person has different suitable auditory stimulation for inducing sleep, we aim to develop a personalized sleep induction system based on mental states.

Machine learning has been used for brain-computer interface to classify various states of the brain [18]–[20]. To monitor the brain, they used an electroencephalogram (EEG), which is the voltage difference between electrodes on the head [21]–[23]. Support vector machine (SVM) was one of the famous machine learning tools for mental state classifications [24]. SVM using the power spectrum of EEG showed 78% accuracy when classifying emotional states between positive and negative emotions [24]. In addition, deep learning like the TinySleepNet was developed for sleep stage classification with 85% accuracy [25]. Although there were more complicated method with higher performance, complicated method demands high computational costs and large memory.

In this study, we developed sleep-inducing system based on metal states which was investigated by the previous study [17]. Our system was designed to consider user's mental states such as sleep quality during the last month, current emotional states, and current cognitive states by the Pittsburgh Sleep Quality Index (PSQI), the Brunel mood scale (BRUMS), and

20xx IEEE. Personal use of this material is permitted. Permission from IEEE must be obtained for all other uses, in any current or future media, including reprinting/republishing this material for advertising or promotional purposes, creating new collective works, for resale or redistribution to servers or lists, or reuse of any copyrighted component of this work in other works. This work was partly supported by Institute for Information & Communications Technology Promotion (IITP) grants funded by the Korea government(MSIT) (No. 2015-0-00185: Development of Intelligent Pattern Recognition Softwares for Ambulatory Brain-Computer Interface, No. 2017-0-00451: Development of BCI based Brain and Cognitive Computing Technology for Recognizing User's Intentions using Deep Learning, No. 2019-0-00079: Artificial Intelligence Graduate School Program, Korea University, and No. 2021-0-02068: Artificial Intelligence Innovation Hub).

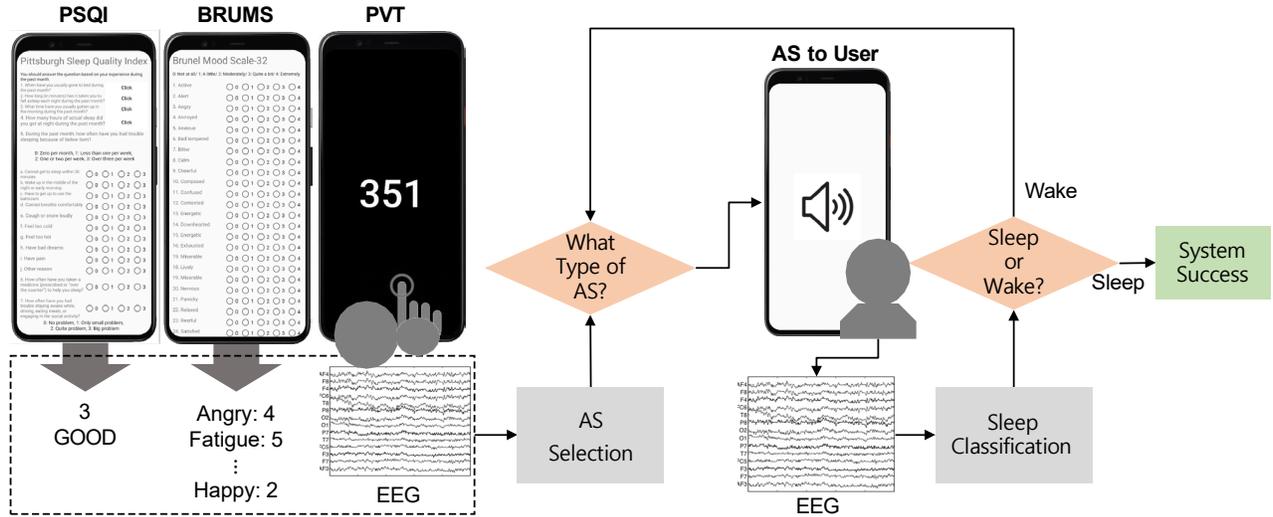

Fig. 1. Overview of Sleep Induction System using Auditory Stimulation (AS).

psychomotor vigilance test (PVT). Their results were exploited for the input of the auditory stimulation selection model. As results of auditory stimulation selection model, the system gave auditory stimulation to users. The EEG during auditory stimulation was used to classify sleep experience. If user fell asleep, the system stopped the auditory stimulation (Fig. 1).

## II. MATERIALS AND METHODS

### A. Participants

We recruited 20 participants to develop sleep induction system. Questionnaires before experiments showed that participants did not take any medication at the time of the experimental session and had no neurological and auditory disease. This study was approved by the Korea University Institutional Board Review (KUIRB-2021-0155-03). Also, written informed consent was obtained from participants.

### B. EEG recording

We obtained EEG signals with 64 channels of Ag/AgCI electrode according to the 10-20 international system using BrainAmp (ActiCap, Brain Products, Germany). The position of reference and ground electrodes were FCz and AFz, respectively. We set the sampling frequency of all electrodes as 1,000 Hz and the kept impedance below 20 kΩ. Since a participant showed high impedance, we excluded the data from the participant when analyzing.

### C. Auditory Stimulation for Sleep Induction

To induce sleep, we investigated five auditory stimulation: sham, RB, BB, WN, and RS. The audio was muted during the sham to investigate the group whose auditory stimulation was not required. We used 512 Hz sound for RB. RB's stimulation interval was 5 sec and it was maintained for 2 sec. We generated the BB using the Gnaural software. BB consisted of 250 Hz and 256 Hz sounds at the left and right ear, respectively [12]. There was a free WN named "pure noise 3" from MC2Method (https://mc2method.org/white-noise/) and a non-licensed RS named "Rain Heavy Loud" from YouTube Studio. Participants choose comfortable sound volume to hear auditory stimulation from 40 dB to 45 dB.

### D. Framework for Sleep Experience Classification

Participants answered the question of whether they fell asleep or not after using the sleep induction system. Our systems classified this sleep experience using real-time EEG and machine learning methods. The input of the system is the 64 channel EEG for 30 sec. We changed the sampling rate from 1000 Hz to 100 Hz and the reference from FCz to Pz. Also, we selected a single EEG from the Oz electrode to obtain Pz-Oz EEG. This is because we pretrained the deep learning model using the Sleep-EDF from PysioNet [26], [27]. There were 153 whole-night Pz-Oz EEG of healthy participants aged 25-101 taking no sleep-related medication. The label was sleep stage decided by the expert sleep technician according to the Rechtschaffen and Kales (R&K) manual [28]. We divided Sleep-EDF data into training and test set. Training set contained 138 participants and test set contained 15 participants. We used cross entropy loss and adam optimizer with weight decay of $10^{-8}$, learning rate of $5 \times 10^{-6}$, and betas $(b_1, b_2) = (0.9, 0.999)$ [29]. The pretrained model estimated the sleep stage of input EEG. Our system collected the data for 10 min and obtained 20 sleep stages. These sleep stages were used to classify sleep experience by fully-connected layer. The fully-connected layer was trained by our 19 participants' data (Fig. 2).

### E. Evaluation of Sleep Induction System

We classified the sleep experience and evaluated our framework with the leave-one-subject-out approach for the generalization performance [30]. Each participant's data were excluded in each training and this excluded participant' data were tested for performance measurement. Since we had 19 participants, we trained 19 models and obtained 19 performances. There was a class imbalance problem and we used not only accuracy but also macro F1 score [31].

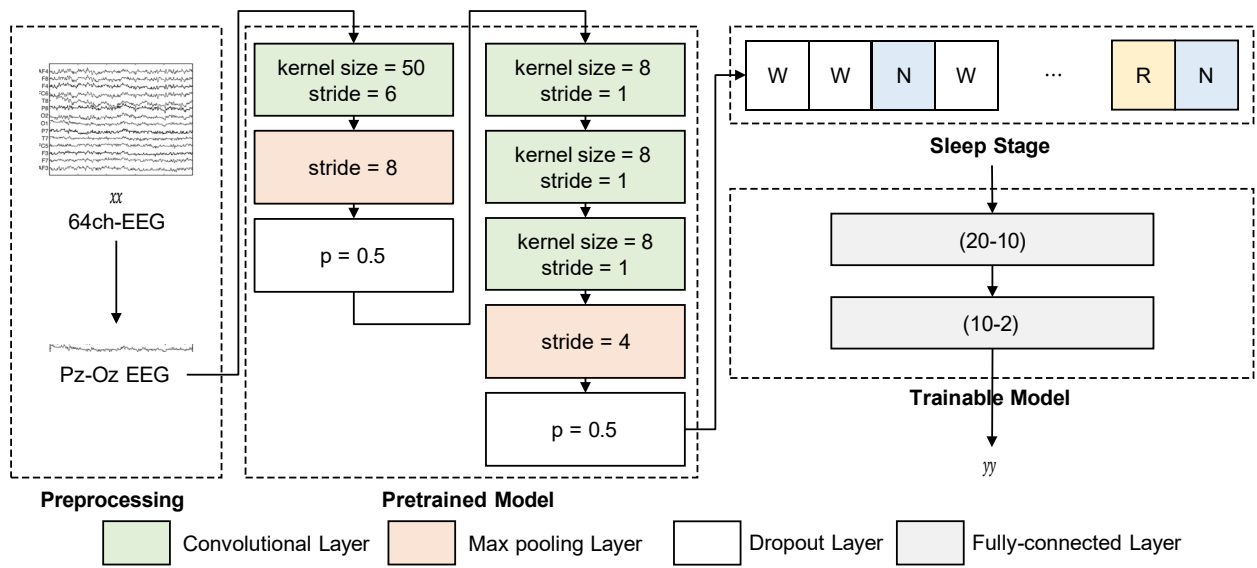

Fig. 2. Framework for sleep experience classification from 64 channel EEG to predicted sleep experience. Sleep stage includes wake (W), non-rapid eye movement sleep (N), and rapid eye movement sleep (R).

TABLE I
SLEEP CLASSIFICATION ACCURACY (ACC) AND MACRO F1 SCORE.

| Subject | ACC | F1 | Subject | ACC | F1 |
|---|---|---|---|---|---|
| 1 | 1.000 | 1.000 | 11 | 0.800 | 0.889 |
| 2 | 1.000 | 1.000 | 12 | 1.000 | 1.000 |
| 3 | 1.000 | 1.000 | 13 | 1.000 | 1.000 |
| 4 | 0.800 | 0.000 | 14 | 1.000 | 1.000 |
| 5 | 0.800 | 0.857 | 15 | 0.800 | 0.857 |
| 6 | 1.000 | 1.000 | 16 | 1.000 | 1.000 |
| 7 | 1.000 | 1.000 | 17 | 1.000 | 1.000 |
| 8 | 1.000 | 1.000 | 18 | 1.000 | 1.000 |
| 9 | 1.000 | 1.000 | 19 | 0.800 | 0.889 |
| 10 | 1.000 | 1.000 | Average | 0.947 | 0.921 |

## III. RESULTS

### A. Pretrained Sleep Stage Classification

Our pretrained sleep stage showed 0.835 macro F1 score and 0.874 accuracy for test set. Training accuracy and macro F1 score were 0.900 and 0.864, respectively.

### B. Sleep Experience Classification

Our framework showed average 0.947 accuracy and 0.921 macro F1 score for sleep experience classification. Except 5 participants, our framework corrected all sleep experience. Since there were 5 samples for a participant, 0.800 accuracy meant that our framework only failed once. A participant who showed 0.800 accuracy and 0.000 F1 score showed the lowest performance. However, except this participant every participants showed high F1 score over 0.800.

### C. Sleep Inducing Effects

When we used sham, RS, RB, BB, and WN, 90% of participants fell asleep. Among 20 participants, only one participant did not fell asleep.

## IV. DISCUSSIONS

In this study, we developed the sleep-inducing system using a previous study. Sleep induction effects of five auditory stimulation based on mental states were investigated by a previous study. Our framework for sleep experience classification showed the high performance to develop a sleep induction system. This framework also showed that the semi-sleep stage from pretrained model was able to be the important key feature for sleep experience classification.

Research showed that RS among autonomous sensory meridian response was able to induce sleep [12]. Nurse in the intensive care unit and coronary care unit was recommended to use WN because it prevents other noise [14], [15]. The monotonous task induced micro sleep [16]. However, these auditory stimulations became disturbing factors to maintain sleep after people fall asleep. Therefore, our system's sleep experience classification framework was useful to stop additional auditory stimulation.

U-time showed a similar idea that the semi-sleep stage from the divided segment enhanced the performance of the ultimate goal [32]. U-time was designed to classify the sleep stage for 30 sec. Before they classified the sleep stage for 30 sec, they classified the sleep stage for each time point. Finally, their segment classifier concluded the sleep stage based on that information. U-time showed high performance like us. Unlike U-time, our framework estimates an explainable semi-sleep stage. U-time's semi-sleep stage was difficult to understand for humans. Our framework's semi-sleep stage help to understand the conclusion of the framework.

Our system should be tested on more participants. Since we had a small sample to support our claims, we will perform experiments with new participants. In addition, we need to compare our semi-sleep stage and real sleep stage. We should investigate these results reproduced in bed at night. Since consciousness research usually used connectivity, connectivity had the potential to increase our system performance [33]–[36].

## CONCLUSION

In conclusion, our sleep-inducing system makes most people fall asleep. Sleep experience classification results are enough to prevent that exceed auditory stimulation that disturb sleep.

## REFERENCES

[1] S.-K. Yeom *et al.*, "Spatio-temporal dynamics of multimodal EEG–fNIRS signals in the loss and recovery of consciousness under sedation using midazolam and propofol," *PLoS One*, vol. 12, no. 11, p. e0187743, 2017.
[2] M. Lee *et al.*, "Network properties in transitions of consciousness during propofol-induced sedation," *Sci. Rep.*, vol. 7, no. 1, pp. 1–13, 2017.
[3] J. J. Pilcher and A. I. Huffcutt, "Effects of sleep deprivation on performance: A meta-analysis," *Sleep*, vol. 19, no. 4, pp. 318–326, 1996.
[4] Y.-S. Kweon, M. Lee, D.-O. Won, and K.-S. Seo, "Prediction of individual propofol requirements based on preoperative EEG signals," in *Int. Winter Conf. Brain-Computer Interface (BCI)*. Gangwon, South Korea, Feb., 2020, pp. 1–5.
[5] K. L. Knutson, K. Spiegel, P. Penev, and E. Van Cauter, "The metabolic consequences of sleep deprivation," *Sleep Med. Rev.*, vol. 11, no. 3, pp. 163–178, 2007.
[6] K.-F. Chung and M.-M. Cheung, "Sleep-wake patterns and sleep disturbance among hong kong chinese adolescents," *Sleep*, vol. 31, no. 2, pp. 185–194, 2008.
[7] W. F. Waters, S. G. Adams Jr, P. Binks, and P. Varnado, "Attention, stress and negative emotion in persistent sleep-onset and sleep-maintenance insomnia," *Sleep*, vol. 16, no. 2, pp. 128–136, 1993.
[8] P. Voitsidis, I. Gliatas, V. Bairachtari, K. Papadopoulou, G. Papageorgiou, E. Parlapani, M. Syngelakis, V. Holeva, and I. Diakogiannis, "Insomnia during the covid-19 pandemic in a greek population," *Psychiatry research*, vol. 289, p. 113076, 2020.
[9] J. Evandt, B. Oftedal, N. Hjertager Krog, P. Nafstad, P. E. Schwarze, and G. Marit Aasvang, "A population-based study on nighttime road traffic noise and insomnia," *Sleep*, vol. 40, no. 2, 2017.
[10] G. Oster, "Auditory beats in the brain," *Sci. Am.*, vol. 229, no. 4, pp. 94–103, 1973.
[11] D. Vernon, G. Peryer, J. Louch, and M. Shaw, "Tracking EEG changes in response to alpha and beta binaural beats," *Int. J. Psychophysiol.*, vol. 93, no. 1, pp. 134–139, 2014.
[12] M. Lee, C.-B. Song, G.-H. Shin, and S.-W. Lee, "Possible effect of binaural beat combined with autonomous sensory meridian response for inducing sleep," *Front. Hum. Neurosci.*, vol. 13, p. 425, 2019.
[13] J. A. Spencer, D. J. Moran, A. Lee, and D. Talbert, "White noise and sleep induction," *Arch. Dis. Child.*, vol. 65, no. 1, pp. 135–137, 1990.
[14] M. L. Stanchina, M. Abu-Hijleh, B. K. Chaudhry, C. C. Carlisle, and R. P. Millman, "The influence of white noise on sleep in subjects exposed to ICU noise," *Sleep Med.*, vol. 6, no. 5, pp. 423–428, 2005.
[15] P. F. Afshar, F. Bahramnezhad, P. Asgari, and M. Shiri, "Effect of white noise on sleep in patients admitted to a coronary care," *J. Caring Sci.*, vol. 5, no. 2, p. 103, 2016.
[16] G. R. Poudel *et al.*, "fMRI correlates of behavioural microsleeps during a continuous visuomotor task," in *Conf. Proc. IEEE Eng. Med. Biol. Soc. (EMBC)*. Minneapolis, USA, Sep., 2009, pp. 2919–2922.
[17] Y.-S. Kweon and G.-H. Shin, "Possibility of sleep induction using auditory stimulation based on mental states," in *2022 10th International Winter Conference on Brain-Computer Interface (BCI)*. IEEE, 2022, pp. 1–4.
[18] D.-O. Won *et al.*, "Motion-based rapid serial visual presentation for gaze-independent brain-computer interfaces," *IEEE Transactions on Neural Systems and Rehabilitation Engineering*, vol. 26, no. 2, pp. 334–343, 2017.
[19] M.-H. Lee, J. Williamson, D.-O. Won, S. Fazli, and S.-W. Lee, "A high performance spelling system based on EEG-EOG signals with visual feedback," *IEEE Trans. Neural Syst. Rehabil. Eng.*, vol. 26, no. 7, pp. 1443–1459, 2018.
[20] H.-I. Suk, S. Fazli, J. Mehnert, K.-R. Müller, and S.-W. Lee, "Predicting bci subject performance using probabilistic spatio-temporal filters," *PloS one*, vol. 9, no. 2, p. e87056, 2014.
[21] O.-Y. Kwon, M.-H. Lee, C. Guan, and S.-W. Lee, "Subject-independent brain–computer interfaces based on deep convolutional neural networks," *IEEE Trans. Neural Netw. Learn. Syst.*, vol. 31, no. 10, pp. 3839–3852, 2019.
[22] J.-H. Jeong, K.-H. Shim, D.-J. Kim, and S.-W. Lee, "Brain-controlled robotic arm system based on multi-directional CNN-BiLSTM network using EEG signals," *IEEE Trans. Neural Syst. Rehabil. Eng.*, vol. 28, no. 5, pp. 1226–1238, 2020.
[23] K.-T. Kim, C. Guan, and S.-W. Lee, "A subject-transfer framework based on single-trial emg analysis using convolutional neural networks," *IEEE Transactions on Neural Systems and Rehabilitation Engineering*, vol. 28, no. 1, pp. 94–103, 2019.
[24] X.-W. Wang, D. Nie, and B.-L. Lu, "Emotional state classification from eeg data using machine learning approach," *Neurocomputing*, vol. 129, pp. 94–106, 2014.
[25] A. Supratak and Y. Guo, "Tinysleepnet: An efficient deep learning model for sleep stage scoring based on raw single-channel eeg," in *2020 42nd Annual International Conference of the IEEE Engineering in Medicine & Biology Society (EMBC)*. IEEE, 2020, pp. 641–644.
[26] B. Kemp, A. H. Zwinderman, B. Tuk, H. A. Kamphuisen, and J. J. Oberye, "Analysis of a sleep-dependent neuronal feedback loop: the slow-wave microcontinuity of the EEG," *IEEE Transactions on Biomedical Engineering*, vol. 47, no. 9, pp. 1185–1194, 2000.
[27] A. L. Goldberger, L. A. Amaral, L. Glass, J. M. Hausdorff, P. C. Ivanov, R. G. Mark, J. E. Mietus, G. B. Moody, C.-K. Peng, and H. E. Stanley, "Physiobank, physiotoolkit, and physionet: components of a new research resource for complex physiologic signals," *circulation*, vol. 101, no. 23, pp. e215–e220, 2000.
[28] D. Moser *et al.*, "Sleep classification according to aasm and rechtschaffen & kales: effects on sleep scoring parameters," *Sleep*, vol. 32, no. 2, pp. 139–149, 2009.
[29] D. P. Kingma and J. Ba, "Adam: A method for stochastic optimization," *arXiv preprint arXiv:1412.6980*, 2014.
[30] M. Esterman, B. J. Tamber-Rosenau, Y.-C. Chiu, and S. Yantis, "Avoiding non-independence in fmri data analysis: leave one subject out," *Neuroimage*, vol. 50, no. 2, pp. 572–576, 2010.
[31] S. Güneş, K. Polat, and Ş. Yosunkaya, "Multi-class f-score feature selection approach to classification of obstructive sleep apnea syndrome," *Expert systems with applications*, vol. 37, no. 2, pp. 998–1004, 2010.
[32] M. Perslev, M. Jensen, S. Darkner, P. J. Jennum, and C. Igel, "U-time: A fully convolutional network for time series segmentation applied to sleep staging," *Advances in Neural Information Processing Systems*, vol. 32, 2019.
[33] Y. Zhang *et al.*, "Strength and similarity guided group-level brain functional network construction for MCI diagnosis," *Pattern Recognit.*, vol. 88, pp. 421–430, 2019.
[34] Y. Zhang, H. Zhang, X. Chen, S.-W. Lee, and D. Shen, "Hybrid high-order functional connectivity networks using resting-state functional MRI for mild cognitive impairment diagnosis," *Sci. Rep.*, vol. 7, no. 1, pp. 1–15, 2017.
[35] M. Lee *et al.*, "Connectivity differences between consciousness and unconsciousness in non-rapid eye movement sleep: A TMS–EEG study," *Sci. Rep.*, vol. 9, no. 1, pp. 1–9, 2019.
[36] K.-H. Thung, P.-T. Yap, E. Adeli, S.-W. Lee, D. Shen, A. D. N. Initiative *et al.*, "Conversion and time-to-conversion predictions of mild cognitive impairment using low-rank affinity pursuit denoising and matrix completion," *Medical image analysis*, vol. 45, pp. 68–82, 2018.